\documentclass[prl,showpacs,superscriptaddress,amsmath,amssymb,reprint]{revtex4-1}
\usepackage{lineno}
\usepackage{graphicx}%
\usepackage{dcolumn}%
\usepackage{bm}%
\usepackage{amsmath,amssymb}%
\usepackage{rotating}%
\usepackage{latexsym}%
\usepackage[usenames]{color}%

\begin{document}

\title{Biopolymer filtration in corrugated nanochannels}

\author{Santtu T. T. Ollila}
\email[e-mail: ]{santtu.ollila@aalto.fi}
\affiliation{COMP Centre of Excellence, Department of Applied Physics, Aalto University School of Science, P.O. Box 11000, FIN-00076 Aalto, Espoo, Finland}
\affiliation{Department of Applied Mathematics, The University of Western Ontario, London, Ontario, Canada N6A 5B8}
\author{Colin Denniston}
\email[e-mail: ]{cdennist@uwo.ca}
\affiliation{Department of Applied Mathematics, The University of Western Ontario, London, Ontario, Canada N6A 5B8}
\author{Mikko Karttunen}
\email[e-mail: ]{mkarttu@gmail.com}
\affiliation{Department of Applied Mathematics, The University of Western Ontario, London, Ontario, Canada N6A 5B8}
\affiliation{Department of Chemistry, University of Waterloo, Waterloo, Ontario, Canada N2L 3G1}
\author{Tapio Ala-Nissila}
\email[e-mail: ]{tapio.ala-nissila@aalto.fi}
\affiliation{COMP Centre of Excellence, Department of Applied Physics, Aalto University School of Science, P.O. Box 11000, FIN-00076 Aalto, Espoo, Finland}
\affiliation{Department of Physics, Brown University, Providence, Rhode Island 02912-1843 USA}
\date{\today}

\begin{abstract}
We examine pressure-driven non-equilibrium transport of linear, circular and star polymers through a nanochannel containing a rectangular pit with full hydrodynamic interactions and thermal fluctuations. We demonstrate that with sufficiently small pressure differences, there is contour length-dependent entropic trapping of the polymer in the pit when the pit and the polymer sizes are compatible. This is due to competition between flow and chain relaxation in the pit, which leads to a non-monotonic dependence of the polymer mobility on its size and should aid in the design of nanofiltration devices based on the polymer size and shape.
\end{abstract}
\pacs{05.40.Jc, 83.10.Rs,47.61.-k}
\maketitle

Biopolymer transport in nano- and microfluidic channels involves competition between external forcing, surface interactions and, possibly, entropic trapping. Careful adjustment of the relative magnitudes between these factors can result in unique transport characteristics such as contour-length/structure based mobility~\cite{SvdHKD06,MRFK11,Persson2009,WM2008,Sakaue2006,Dorfman2010}, helical coiling~\cite{CWG2012}, pit-to-pit hopping in pressure-driven flow~\cite{BRS09,MRFK11,ZdePG2012}, and electrophoretic mobility differences for long DNA molecules~\cite{HTC1999,HC2000}. Electric field-driven separation techniques have also been optimized for DNA~\cite{HC2002}. Mikkelsen {\it et al.} recently demonstrated pressure-driven DNA to exhibit clear mobility differentiation both as a function of the degree of polymerization, $N$, and internal structure (linear and ring with the same $N$)~\cite{MRFK11}. Their experimental arrangement consisted of an array of troughs in which the chain was transiently trapped.

\begin{figure}[b]
\centerline{\includegraphics[width=0.9\columnwidth]{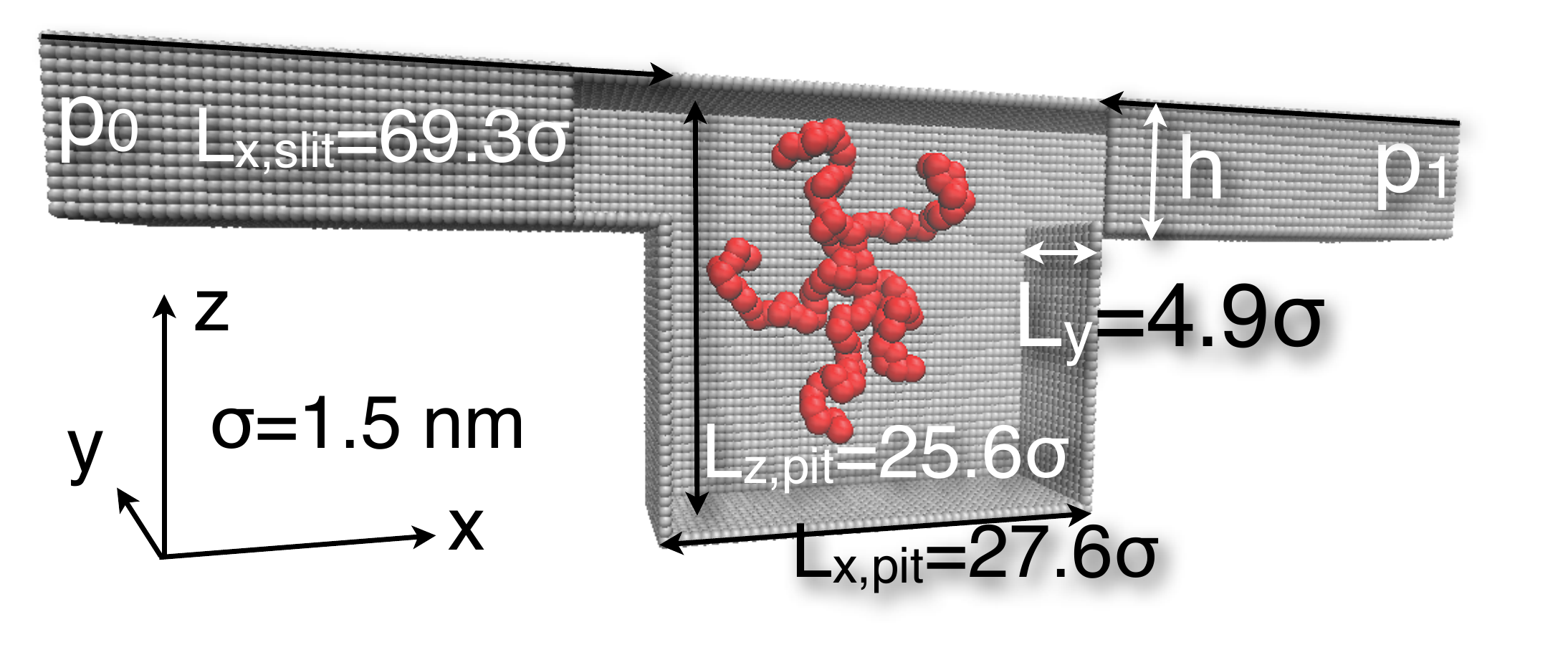}}
\caption{Flow is in the +$x$ direction due to a pressure differential $\Delta p = p_0 - p_1 > 0$ between the ends. The cutout reveals the polymer and the interior (created using VMD~\cite{HDS96}).}
\label{fig:fig1}
\end{figure}

In this work, we show how a nanochannel with sidewalls and alternating sections of square narrows (``slits") and rectangular pits can be used for pressure-driven polymer filtration (see Fig.~\ref{fig:fig1}). The change in the channel's cross section facilitates swelling of the chain in a time frame dictated by a pressure differential and thermal fluctuations, and it leads to non-trivial effects on chain mobility as a function of $N$ and chain structure. We also test the validity of the threshold of minimal momentum flux needed for a polymer to enter a narrow pore~\cite{SRdeGB-W05} in our geometry. The threshold was recently found to be accurate in simulations under body force driven flow~\cite{MUABY09}. However, a body force and a pressure differential are not equivalent in a channel of varying cross-section as the pressure drop per unit length decreases as the cross-sectional area increases. We demonstrate that the trapping effect here is due to competition between flow and chain relaxation in the pit, leading to non-monotonic dependence of the polymer mobility on the polymer size.

Polymer dynamics in confinement is affected by the relative size of the channel and the polymer, the latter of which is characterized in free space by the isotropic radius of gyration, $R_g$. It is defined via $R_g^2 = \langle \sum_{n=1}^N ({\bf r}_n - {\bf r}_\mathrm{cm})^2\rangle/N$, where ${\bf r}_n$ is the $n^\mathrm{th}$ monomer's position and ${\bf r}_\mathrm{cm}$ is the polymer's center of mass.  Applying a pressure differential $\Delta p = p_0 - p_1 > 0$ between the ends of the system in Fig.~\ref{fig:fig1} results in a polymer mobility defined as
\begin{equation}
\label{eq:mobility}
\mu = \langle v_x \rangle / \Delta p,
\end{equation}where $\langle v_x \rangle$ is the mean (weighted spatial average over the system) velocity of the polymer in the $x$ direction.

The mobility depends on the relative magnitude of $\Delta p$ and thermal fluctuations, which determine how quickly the polymer can move perpendicular to flow streamlines. Unlike the electrophoretic mobilities in Ref.~\cite{Sakaue2006} that were determined theoretically based on an equilibrium barrier crossing approach via the Fokker-Planck equation, we are able to measure mobilities in the presence of non-equilibrium flow effects, such as the polymer blocking flow in the slit, as our methodology is not restricted to near-equilibrium. Also, driving due to an electric field in the slit only~\cite{Sakaue2006} is distinctly different from our case, where {\it e.g.}, there is always an upward force in the pit pushing the chain toward the slit and the flow can unravel the chain through tumbling affecting the escape dynamics.

We change the ratio of the competing factors contributing to $\mu$ by varying $\Delta p$ that results in a mean solvent momentum flux $\bar{j_x}$ we measure over a cross section $A$ of the channel and average over time:
\begin{equation}
\bar{j_x}A = \Bigl\langle \int_A \rho({\bf r}) u_x({\bf r}) \,\mathrm{d} A \Bigr\rangle_t,
 \end{equation}where the local mass density and the components of fluid velocity are denoted by $\rho({\bf r})$ and $u_\alpha ({\bf r})$ with $\alpha\,\in\,\lbrace x,y,z\rbrace$.
The pit-to-slit transition has been examined theoretically by Sakaue {\it et al.} using the blob theory~\cite{SRdeGB-W05}. Their result is that the momentum flux is able to make a flexible polymer enter a narrows smaller than $R_g$ providing the flux exceeds a threshold specific to the process. We cast the threshold here in terms of the fluid momentum density, $j_c$, at the pit-to-slit interface as
 \begin{equation}
 \label{eq:threshold}
 j_c A \simeq k_\mathrm{B} T / \nu,
 \end{equation}where $k_\mathrm{B}$ is Boltzmann's constant, $T$ the temperature and $\nu$ the kinematic viscosity of the solvent. We emphasize that Eq.~(\ref{eq:threshold}) is independent of pore details and it is obtained by assuming a very long polymer and the process to be slow enough to be describable by a free energy that exhibits a barrier of entry as a function of the partial length of the pore occupied by the polymer.

In the absence of bias, the equilibrium barrier of entry is dictated by the free energy difference, $\Delta F$, between the sections. Experiments can have a large $\Delta F$ by imposing Odijk confinement in the narrows (of height comparable to the chain persistence length $l_p$) and de Gennes confinement in the pit ($R_g > \mathrm{height} > l_p$). Such a large contrast is unfeasible for the coarse-grained model we use here. However, we see sufficient contrast to clearly illustrate the important factors driving the dynamics.

Our system is a channel with alternating segments of slits and pits (Fig.~\ref{fig:fig1}). The slit has a square cross section of size $L_y \times h = 4.9\,\sigma \times 4.9\,\sigma$ and the pit has a rectangular cross section $L_y \times L_{z,\mathrm{pit}} = 4.9\,\sigma \times 25.6\,\sigma$ (this is the space available for the polymer with the monomer-wall interaction cutoff subtracted). The pit and slit lengths in the $x$ direction are $L_{x,\mathrm{pit}}=27.6\,\sigma$ and $69.3\,\sigma$, respectively. $\sigma=1.5\,\mathrm{nm}$ defines the length scale in the $12$-$6$ truncated and shifted Lennard-Jones potential (LJ), $U_\mathrm{LJ}=4\epsilon((\sigma/r)^{12}-(\sigma/r)^6+1/4)$, where the energy scale is $\epsilon=k_\mathrm{B} T$ and $T=300\,\mathrm{K}$. The polymer is modeled by joining consecutive beads with a FENE bond~\cite{GK86} with a maximum length of $R_0=1.5\sigma$ and a spring constant of $k=32.6\epsilon\sigma^{-2}$. The excluded-volume effect is realized through $U_\mathrm{LJ}$ between all monomer pairs. The channel walls are made out of LJ beads with $\epsilon$ as above and the monomer-wall distance parameter set to $\sigma_\mathrm{mw}=1.87\,\sigma$. We follow chains consisting of $N\!=\!16$ to $96$ beads moving in the channel with a skew-periodic boundary condition (sPBC) splitting the narrows into halves of equal length.

A fluctuating Navier-Stokes solver is responsible for both the Brownian and the advective motion the chain undergoes. The solver is based on the lattice-Boltzmann equation and equilibrium velocity fluctuations $\langle u_\alpha^2 \rangle \sim T$ in the fluid velocity arise from a random component in the fluid stress tensor along with corresponding fluctuations in higher moments~\cite{ODKA11}. The advective flow is due to a pressure differential in the $x$ direction set as an sPBC through $\Delta p = (\rho_0 - \rho_1)v^2_s$, where $\rho_i$ are the set mean densities at the ends (see Fig.~\ref{fig:fig1}), and the equation of state $p=\rho v^2_s$ ($v_s$: speed of sound). An important physical distinction is that the resulting pressure drop in the pit is less than that in the slit while the mass flux stays constant: unlike a body force, a pressure BC allows the fluid to slow down when the channel's cross sectional area increases. The solvent couples to the monomers through a hydrodynamically consistent scheme and provides a well-defined hydrodynamic radius for the monomers~\cite{ODKA11,OSAD12,MOD13,DMO13}. There is a no-slip BC based on the mid-grid bounceback rule on the walls. The kinematic viscosity in our simulations is $\nu = 1403\,\mathrm{nm}^2\,\mathrm{ns}^{-1}$, which is $1.4$ times that of water and the resolution of the LB mesh is $\Delta x = 1.0\,\mathrm{nm}$.

To compute $\mu$, we analyzed time series of the $x$ component of the chain's CM coordinate, $x_\mathrm{cm}$. We adjust the balance between entropic trapping and advection by varying $\Delta \rho$ between the ends of the system relative to the mean density from $q \equiv \Delta \rho/\langle \rho \rangle = 7.6 \times 10^{-4}$ to $3.8 \times 10^{-3}$. In the corresponding square channel (in the absence of the pit), this range of $\Delta \rho$ would result in a flow profile whose mean velocity $\langle u \rangle$ varies between $\langle u \rangle/(\sqrt{k_\mathrm{B}T/(\langle \rho \rangle \Delta x^3)})=7.6 \times 10^{-4}$ and $3.8 \times 10^{-3}$. The mean bias is chosen sufficiently small relative to velocity fluctuations in the solvent to allow for spontaneous particle migration across streamlines. This has to be balanced by an increase in runtime limiting us from going to weaker driving which would give greater mobility differentiation. The Reynolds number, $\mathrm{Re} \equiv L_y \langle v_x \rangle/\nu$, varies between $0.002$ and $0.016$. We also define the P\'eclet number $\mathrm{Pe} \equiv L_y \langle v_x \rangle/D$, where $D$ is the diffusion coefficient of the chain in a corresponding level of slitlike confinement. We find $1 \leq \mathrm{Pe} \leq 25$ for the range of $\Delta p$ considered.
\begin{figure}
\centerline{\includegraphics[width=.82\columnwidth]{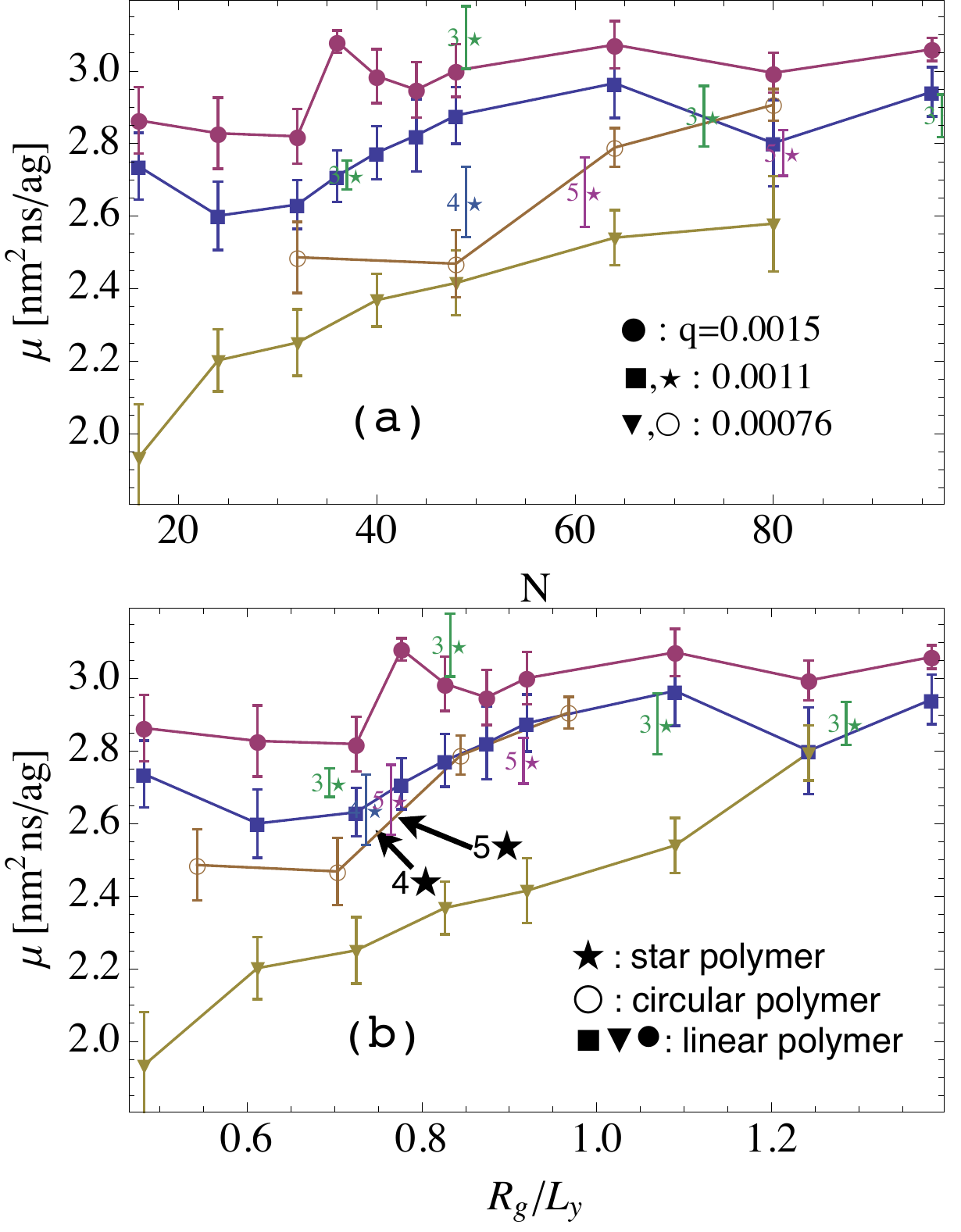}}
\caption{Polymer mobility $\mu$ as a function of (a) $N$ and (b) the relative chain size ($q \equiv \Delta \rho/\langle \rho \rangle=\Delta p/\langle p \rangle$).}
\label{fig:fig2}
\end{figure}

We have determined $\mu(N)$ in the case of linear, star and circular polymers. Mobility allows us to characterize polymer movement at different $\Delta p$ on an equal footing. This measurement is interesting since, if $\mu$ were to depend on $N$, our system could be applied to polymer filtration based on length or structure. Figure~\ref{fig:fig2} shows $\mu(N)$ for different $\Delta p$. The $\star$ indicate $f$-branch star polymers of $N=f\,N_f+1$ monomers ($12 \leq N_f \leq 32$), the $\circ$ are circular, and the other polymers are linear. We observed no traces of entropic trapping at $q = 0.0038$ (data not shown) as $\Delta p$ dominates the dynamics. This leads to an $N$-independent value of $\mu$ as everything is flushed through the system. However, at $q = 0.0011$ ($\blacksquare$), the situation becomes more interesting as $\mu(N)$ dips at $N=24$, then increases monotonically up to $N=64$. The result is significant as the error in the averages is smaller than $\mu(64)-\mu(24)$. Most significantly at the lowest $\Delta p$, there is a clear increase of $\mu(80)/\mu(16)\approx 1.45$ in mobility ($\blacktriangledown$).

The non-monotonicity for the $\blacksquare$ case illustrates the different driving forces on the chains: first, the time available to relax in the pit, and second, the entropic resistance to exiting the pit. Both effects are impacted by the variation in flow speed (the driving force) present in pressure driven flow (as opposed to a constant body force).  The streamline velocity in the $x$ direction is higher at the top of the pit than at the bottom.  Long chains have higher mobility as part of the chain is always in the faster moving streamlines at the top.  Once any part of a chain reaches the exit, it will experience an increasing force, as it is blocking flow, until it is pushed out (a key difference between pressure driving and constant body force).  Given time for thermal motion to act, short chains can escape the faster streamlines and thus spend more time in the depths of the pit before wandering to the exit. Intermediate length chains expand out of the fast streamlines upon entering the pit which acts as a partial brake on their $x$ velocity, potentially giving them enough time to migrate completely off the faster streamlines and into the depth of the pit. So, for weak driving these effects are cooperative and both favor short chains loitering in the pit. However, for intermediate driving the short chains do not have enough time to migrate out of the fast streamlines and so exit rapidly.  However, the ``brake'' effect of part of the chain dipping into the slow moving flow for intermediate chain lengths can give them enough time to get out of the fast moving streamlines entirely, thus making their mobility lower than both short and long chains'.  
Reducing $\Delta p$ increases the time available for chains to migrate off the fast streamlines and relax into the pit, which, in turn, changes the $\mu$ vs. $N$ data.  Equilibrium barrier crossing arguments become more valid as $\Delta p \rightarrow 0$.
\begin{figure}
\centerline{\includegraphics[width=0.825\columnwidth]{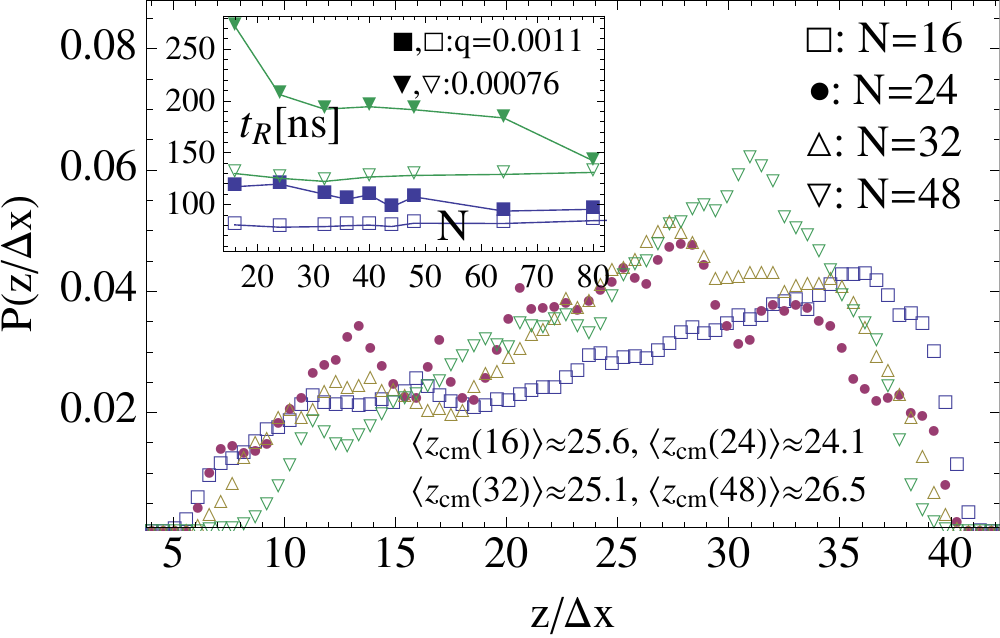}}
\caption{Probability distribution of $z_\mathrm{cm}$ in the pit for $q = 0.0011$ ($\blacksquare$ in Fig.~\ref{fig:fig2}). Note $N=24$ has a peak at the bottom of the pit and it also spends most time in the pit. Inset: residence time $t_R(N)$ in the pit (solid) and slit (hollow symbols). $t_R$ increases especially for small $N$ in the pit as $q$ decreases.}
\label{fig:Pzcm}
\end{figure}

These aspects are visible in Fig.~\ref{fig:Pzcm} in the probability distributions of the $z$ component of the chains' CM coordinate, $z_\mathrm{cm}(N)$, in the pit for $16\leq N\leq 48$ even at $q = 0.0011$. The distributions are consistent with the corresponding $\blacksquare$ in Fig.~\ref{fig:fig2} as the averages $\langle z_\mathrm{cm} \rangle=\int_{z=0}^{z=L_z} z P(z)\,\mathrm{d}z$ are ordered as $\langle z_\mathrm{cm}(24) \rangle < \langle z_\mathrm{cm}(32) \rangle  < \langle z_\mathrm{cm}(16) \rangle < \langle z_\mathrm{cm}(48) \rangle$ and so are the corresponding mobilities. The inset of Fig.~\ref{fig:Pzcm} shows chain residence times in the pit (solid) and slit (hollow symbols) vs. $N$. Chains slow down in the pit and the $N$ dependence strengthens for $q=0.00076$. This change increases the time spent in the pit (more than that in the slit) during which short chains have a larger probability ($D_\mathrm{cm} \sim N^{-\nu}$) and ability ($R_g \sim N^{\nu}$) to move deeper into the pit.

The significance of Fig.~\ref{fig:fig2} can be understood by relating it to experiments in terms of the degree of confinement (C) and the geometry of the system. Stein {\it et al.}~\cite{SvdHKD06} have determined experimentally the mobility of double-stranded DNA in pressure-driven flow in slitlike silica channels of height from $d_\mathrm{min}\approx 0.2\,\mu\mathrm{m}$ to $d_\mathrm{max}\approx 3.8\,\mu\mathrm{m}$. They found the mobility of $4\,\mu\mathrm{m}$ to $22\,\mu\mathrm{m}$-long DNA to be the same at $d_\mathrm{min}$ (due to identical dispersion characteristics and concentration profiles in the channel), but the mobility of the shortest chain dropped by $11\%$ compared to the longest at $d_\mathrm{max}$. They related the drop to an $N$-dependent density distribution stemming from contour-length dependent Taylor dispersion only seen in wide channels corresponding to $C = R_g/d \lesssim1.5$~\cite{SvdHKD06}. Here, we see a similar difference in $\mu(N)$ for a smaller ratio between contour lengths and the same level of $C$ in the slit at $q=0.0015$: $\mu(64)$$/\mu(24) \approx 1.14$ and $C=R_g(24)$$/L_y \approx 0.62,\ldots,R_g(64)/L_y \approx 1.1$. At $q = 0.00076$, the ratio stays roughly the same, but the monotonic increase in $\mu$ extends from $N\!=\!16$ to $80$.

The mobility at larger $\Delta p$ is roughly constant for $N\geq 80$ independent of structure (number of branches). For smaller $N$, Fig.~\ref{fig:fig2}(a) shows star polymers $(f,N_f)\!=\!(3,16)$ and $(4,12)$ move at very different velocities. However, as the mobility is plotted as a function of $R_g/L_y$ in Fig.~\ref{fig:fig2}(b), most star polymers with $N<64$ or $R_g/L_y \lesssim 1.0$ have mobilities very close to those of linear chains of the same size.   However, circular chains enter the narrows of our system as a hairpin through pinching (by definition), whereas linear chains are required to fold to do the same. This kind of dynamics suggests that circular chains should have higher mobilities than their linear counterparts (in the hairpin there are always two parallel strands in the slit, hence roughly twice the driving force experienced by single-file motion).  This is indeed what we find in Fig. \ref{fig:fig2}.  These observations support the finding in Ref.~\cite{MRFK11} that artificial nanotopographies can lead to technologically significant mobility differentiation devices based on internal chain structure. 

The regime of validity of our simulation methodology extends at least to $C\approx 3$ within the de Gennes regime ($L_y/l_p \gg C > 1$)~\cite{ODKA12}. Our simulations indicate that a nanopit array (height contrast $L_{z,\mathrm{pit}}/h \approx 5.2$) with sidewalls leads at least to the same level of mobility differentiation within the de Gennes regime as a slit. However, Mikkelsen {\it et al.}~\cite{MRFK11} observed mobility ratios up to $100$ between linear DNA chains of lengths $165.6\,\mathrm{kbp}$ and $48.5\,\mathrm{kbp}$ for $L_{z,\mathrm{pit}}/h = 3$ in an array of troughs. Their work was distinctly different from ours in two ways. First, their slit height was comparable to the chain's persistence length, which corresponds to the Odijk regime~\cite{Odijk1983}. Second, their pits were long troughs, which impose no confinement in the $y$ direction of Fig.~\ref{fig:fig1}. 

We discuss the second aspect first. Mikkelsen {\it et al.} found both linear and circular chains to escape the pit either by its end (circular chain rolls out at both ends) threading the pit-to-slit interface first ("sidewinder" mode) or by the middle section of the chain forming a hairpin and thus initiating the propagation ("tumbleweed")~\cite{MRFK11}. We have seen both modes in simulations without sidewalls at $q = 0.0015$. Thus, the modes are determined strongly by the value of $C$ in the $y$ direction. However, the degree of mobility differentiation appears to be dominated by the slit height, which we discuss next.

That $\mu(N=48)$ through $\mu(96)$ is roughly constant down to $q = 0.0011$ suggests that even $\bar{j}(0.0011) > j_c$ here, assuming $j_c$ is independent of $N$. However, the ratio of $h$ to the chain's mean diameter, $h/(2 R_g)$, varies from $0.9$ ($N=16$) to $0.3$ ($N=96$), making it debatable whether the slit in our system is small enough for Eq.~(\ref{eq:threshold}) to hold. Despite the prefactor in Eq.~(\ref{eq:threshold}) being unknown~\cite{SRdeGB-W05,MUABY09}, it is interesting to compare the prediction of Eq.~(\ref{eq:threshold}) for the threshold to the averages at different $\Delta p$ in our simulations. We have $\bar{j}(q=0.00076)/ j_c=0.5$, $\bar{j}(0.0011)/ j_c=0.8$ and $\bar{j}(0.0015)/ j_c=1.1$. These numbers and the constant value of $\mu(N=48-96)$ for $q=0.0011$ suggest the prefactor is less than $0.5$ for our system. The weakest $\Delta p$ ($\blacktriangledown$ in Fig.~\ref{fig:fig2}) may be close to a threshold for our system as the dependence $\mu(N)$ strengthens compared to stronger driving. 

Another adjustable feature in the channel design is the pit depth $L_{z,\mathrm{pit}}$, which in this work was optimized to provide a maximal downward solvent flux in the upstream half of the pit in order to prolong the trapping time of the polymer. Deeper pits do not result in additional downward flux, but they do provide more space for the polymer to take up. Thus, increasing $L_{z,\mathrm{pit}}$ could translate to larger mobility differences as a function of $N$ or possibilities for intermittent chain extraction from the system, and should be investigated in future work.
\begin{figure}
\centerline{\includegraphics[width=.8\columnwidth]{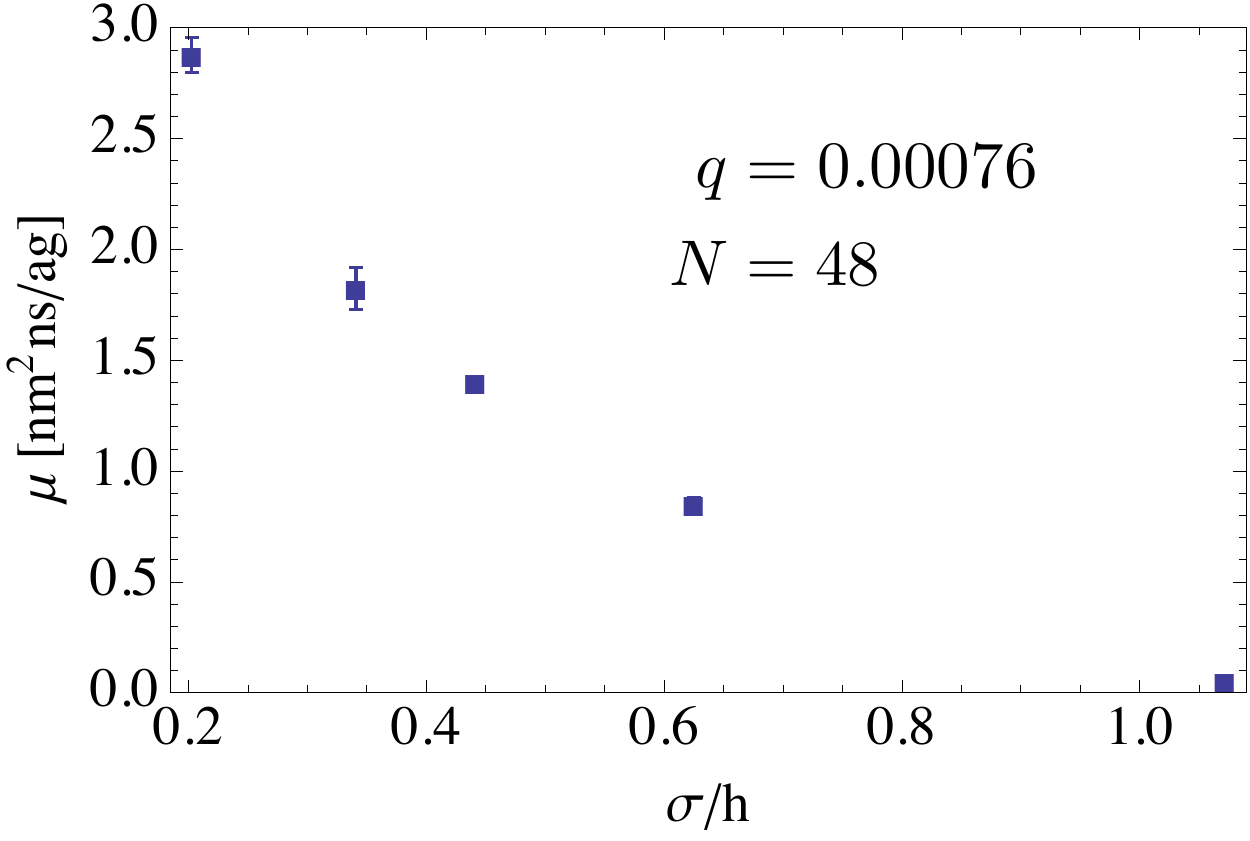}}
\caption{Polymer mobility versus the inverse of the slit height.}
\label{fig:fig4}
\end{figure}

The free energy difference between the slit and the pit can be increased by reducing $h$. The effect of $h$ on $\mu$ shown in Fig.~\ref{fig:fig4} gives a decrease of roughly an order of magnitude. In this case, the slit height decreases close to that in studies of polymer translocation~\cite{MHS2013,Dekker2007}. Unfortunately, a shallower slit would have made runtimes too long, but it would have given a larger mobility difference.

{\it Conclusions --}We have studied the passing of a polymer through a nanoarray in pressure-driven flow, and demonstrated entropic trapping for $\mathrm{Pe}<10$. The observed non-monotonic dependence of the polymer mobility on its size can be greatly amplified in experiments, which can access much smaller pressure differentials than simulations. The threshold value of $\Delta p$, where the trapping emerges, is interesting in its own right as it highlights the role of the relevant system parameters, including the pit size. Our results demonstrate the feasibility of building simple but highly selective nanofiltering devices, which can differentiate polymers based on their size and shape.
\begin{acknowledgments}
This work was supported by the Academy of Finland through its COMP CoE grant No. 251748, EXPECTS grant from Aalto, and the Natural Science and Engineering Council of Canada. Computational resources were provided by Aalto Science-IT project, SharcNet, and Compute/Calcul Canada.  This research was supported in part by the National Science Foundation under Grant no. NSF PHY11-25915.
\end{acknowledgments}



\begin{thebibliography}{55}
\bibitem{SvdHKD06}
D. Stein, F. H. J. van der Heyden, W. J. A. Koopmans, and C. Dekker, {\it P Natl Acad Sci USA} {\bf 103} 15853 (2006).

\bibitem{MRFK11}
M. B. Mikkelsen, W. Reisner, H. Flyvbjerg, and A. Kristensen, {\it Nano Lett.} {\bf 11} 1598 (2011).

\bibitem{WM2008}
C. T. A. Wong and M. Muthukumar, {\it Biophys. J.} {\bf 95} 3619 (2008).

\bibitem{Sakaue2006}
T. Sakaue, {\it Eur. Phys. J.} E {\bf 19}  477 (2006).

\bibitem{Persson2009}
F. Persson, P. Utko, W. Reisner, N.B. Larsen, and A. Kristensen, Nano. Lett. {\bf 9}, 1382 (2009).

\bibitem{Dorfman2010}
K. Dorfman, {\it Rev. Mod. Phys.} {\bf 82} 2903 (2010).

\bibitem{CWG2012}
R. Chelakkot, R. G. Winkler, and G. Gompper, {\it Phys. Rev. Lett.} {\bf 109} 178101 (2012).

\bibitem{ZdePG2012}
Y. Zhang, J. J. de Pablo, and M. D. Graham, {\it J. Chem. Phys.} {\bf 136} 014901 (2012).

\bibitem{BRS09}
J. T. Del Bonis-O'Donnell, W. Reisner, and D. Stein, {\it New J. of Phys.} {\bf 11} 075032 (2009).

\bibitem{HTC1999}
J. Han, S. W. Turner, and H. G. Craighead, {\it Phys. Rev. Lett.} {\bf 83} 1688 (1999).
\bibitem{HC2000}
J. Han and H. G. Craighead, {\it Science} {\bf 288} 1026 (2000).
\bibitem{HC2002}
J. Han and H. G. Craighead, {\it Anal. Chem.} {\bf 74} 394 (2002).

\bibitem{SRdeGB-W05}
T. Sakaue, E. Raphael, P.-G. de Gennes, and F. Brochard-Wyart {\it Europhys. Lett.}, {\bf 72} 83 (2005).
\bibitem{MUABY09}
A. P. Markesteijn, O. Berk Usta, Issam Ali, Anna C. Balazs, and J. M. Yeomans, {\it Soft Matter} {\bf 5} 4575 (2009).

\bibitem{ODKA11}
S. T. T. Ollila, C. Denniston, M. Karttunen and T. Ala-Nissila, {\it J. Chem. Phys.} {\bf 134} 064902 (2011).
\bibitem{OSAD12}
S. T. T. Ollila, C. J. Smith, T. Ala-Nissila, and C. Denniston, {\it Multiscale Model. Simul.} {\bf 11} 213 (2013).

\bibitem{MOD13}
F. E. Mackay, S. T. T. Ollila, and C. Denniston, {\it Comp. Phys. Comm.} {\bf 184} 2021 (2013).
\bibitem{DMO13}
C. Denniston, F. E. Mackay, and S. T. T. Ollila, {\it NIC Series} {\bf 46}  232 (2013).

\bibitem{ODKA12}
S. T. T. Ollila, C. Denniston, M. Karttunen and T. Ala-Nissila, {\it Soft Matter} {\bf 9} 3478 (2013).

\bibitem{Odijk1983}
T. Odijk, {\it Macromolecules} {\bf 16} 1340 (1983).

\bibitem{HDS96}
W. Humphrey, A. Dalke and K. Schulten, {\it J. Molec. Graphics}  {\bf 14} 33 (1996).

\bibitem{GK86}
G. S. Grest and K. Kremer, {\it Phys. Rev.} A {\bf 33} 3628 (1986).

\bibitem{MHS2013}
M. Mihovilovic, N. Hagerty, and D. Stein, {\it Phys. Rev. Lett.} {\bf 110} 028102 (2013).

\bibitem{Dekker2007}
C. Dekker, {\it Nature Nanotech.} {\bf 2} 209 (2007).

\end{thebibliography}
\end{document}